# Mechanisms in Slide Electrification of Liquid and Frozen Drops on Hydrophobic Surfaces

Rutvik Lathia[1,*]; Benjamin Leibauer[1]; Aaron D. Ratschow[1]; Werner Steffen[1] and Hans-Jürgen Butt[1,*]

[1] Max Planck Institute for Polymer Research, Ackermannweg 10, 55128 Mainz, Germany

*Corresponding authors: lathiar@mpip-mainz.mpg.de; butt@mpip-mainz.mpg.de

## Abstract

The microscopic and fundamental origin of slide electrification, where droplets of water move across insulating surfaces accumulating and depositing electrical charges, is still debated. Charge transfer is often attributed to ion transfer at the receding contact line. However, it is still unclear whether ion transfer alone can fully account for the observed charge separation. We examined slide electrification of two polar, self-ionizing liquids (water, formamide) and two non-polar liquids (diiodomethane, bromonaphthalene). By cooling below the melting temperature, we were able to compare this process to tribocharging of the respective frozen components. Despite reduced ion mobility at sub-freezing temperatures, the frozen polar compounds continue to accumulate significant charge. Non-polar liquids exhibit lower charging (<25% of polar liquids) and nearly identical charging behaviour in both their liquid and frozen phases on five different substrates. Since non-polar liquids contain few free ions, these observations indicate an alternative charging mechanism, which could be electron transfer. Our findings suggest that slide electrification operates through at least two mechanisms, with the dominant charge transfer pathway shifting between ions and electron transfer depending on the electronegativity, phase, and temperature.

Water droplets sliding over insulating and hydrophobic surfaces can accumulate electric charge. Consequently, the solid surfaces underneath a droplet acquire opposite charges, a process referred to as slide[1,2] or contact electrification, analogous to triboelectricity between solids.[3] It is not clear how charges are transferred, given that bringing a unit charge from the solid-water interface to the free solid surface is highly energetically unfavourable. Unlike solid tribocharging, where high shear stresses easily break chemical bonds at asperities,[5,6] the contact line of a drop is soft. While the charging of droplets is recognized for over a century,[7,8] quantitative experiments have been scarce until recent times. In the past decade protocols have emerged to quantify this charging of sliding drops.[9-12] Understanding the mechanisms behind water droplet electrification is essential to understanding dynamic wetting and holds significant implications in many areas related to condensation, energy generation, printing, and desalination.[13-15]

Despite extensive efforts, the microscopic origin of charge transfer in slide electrification remains debated.[16-18] For liquid water, the prevailing view attributes charging primarily to surface charges, which form spontaneously at the water-solid interface. Usually, the bound surface charges are formed by dissociation of surface groups such as amino groups[19] or the adsorption of ions that remain on the solid surface at the receding contact line. This initial (or primary) charge separation is driven by chemical energy. As the contact line of the drop recedes, the counterions in the diffuse part of the electric double layer accumulate in the sliding drop while the surface charge remains on the dewetted surface.[2,20,21] For hydrophobic surfaces there is still no consensus of how surface charges are formed.[16-19,22] Zeta potential measurements,[23] potentiometric titration[24,25] and surface force experiments[26] demonstrate that they are usually negatively charged. These observations suggest that an enrichment of $OH^-$ at the interfaces leads to surface charging. As an additional mechanism to ion transfer, electron transfer has been proposed, wherein direct electronic exchange due to electron cloud overlap of the solid–liquid interface contributes to surface charging.[16-18,22] Electron transfer is commonly considered as the dominant charge transfer mechanism in contact electrification for solid–solid contact-pairs of metals and semiconductors. For insulators, and in particular for polymers, electron transfer is not convincing.[27-29] Charge transfer in contact electrification of insulators is still debated.

It is well known that ice-vapor and ice-solid interfaces of water are often charged.[30-32] In ice, the main charge carriers are protons (or $H_3O^+$) and $OH^-$, but not electrons.[4,30] Tribocharging of solid surfaces by ice can lead to charge densities above 1 $mC/m^2$. The effect has even been utilized to build triboelectric nanogenerators.[33,34] For ice, the formation of the diffuse part of an electric double layer will be supressed. Although a liquid-like interfacial layer is often reported at ice surfaces and solid–ice interfaces, its thickness under sub-freezing conditions is generally considered to be much smaller than a typical electric double-layer thickness. Furthermore, it is expected to decrease with decreasing temperature to below 1 nm.[35-37] For this reason one would expect that charge separation occurring between a solid and ice is different from slide electrification between a solid and liquid water.

Here we ask whether the charging observed in liquids and their frozen counterparts is fully captured by an ion-transfer-based description or whether and which additional mechanisms contribute. To elucidate the fundamental process(es) of charge transfer, we studied the tribocharging and the slide electrification of frozen and liquid water, formamide (both polar), diiodomethane and 1-bromonaphthalene (both non-polar). Formamide is another polar self-ionizing liquid ($2HCONH_2 \rightleftharpoons HCONH_3^+ + HCONH^-$, $pK_a = 16.8$)[38]. Unlike water and formamide, non-polar liquids do not contain free ions because the dissociation of chemical groups or dissolution of salt is energetically too costly. Accordingly, to our knowledge, substantial slide electrification has only been observed for polar liquids.[39] He and Darhuber observed that, for a series of liquids, the generation of charges is at least ten times lower for liquids with a dielectric permittivity $\varepsilon < 30$ than for water.[40] Similar observations were found in other reports as well.[41,42] The finite conductivity caused by contamination in non-polar liquids, is often responsible for the electrification.[43]

We performed the experiments on five chemically different surfaces. Using a tilted plate setup with capacitive current detection[12], we quantified charge accumulation under controlled temperature conditions. Our results show that the frozen state of polar liquids, despite reduced ionic mobility, continues to charge significantly, with enhanced charging near the melting transition. Nonpolar liquids, in both liquid and frozen phases, displayed consistent charge deposition indicating an additional mechanism.

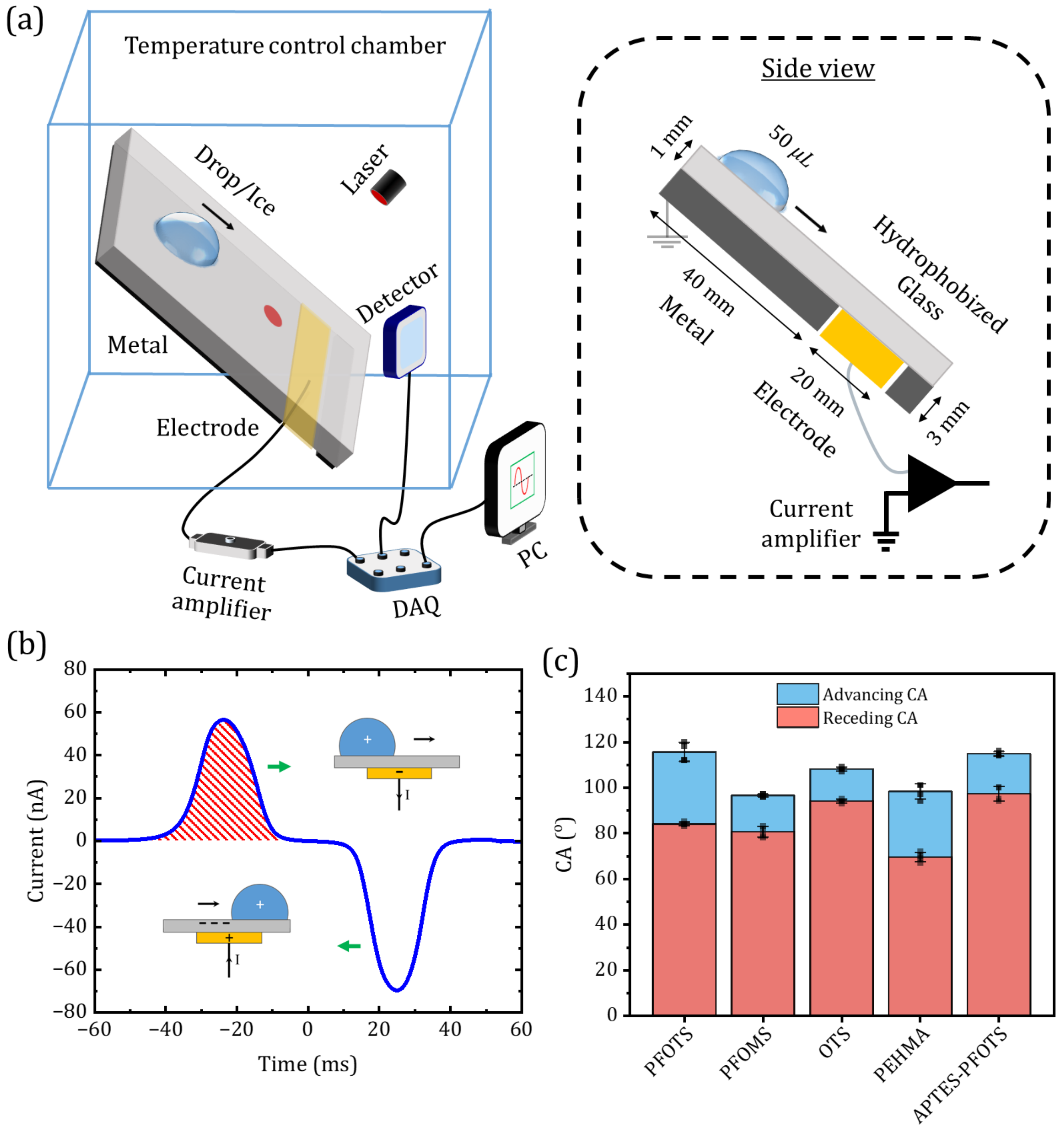

*Fig. 1: (a) Setup for measuring the charge on droplets and ice. The setup consists of a hydrophobized glass plate inclined by 50º with a copper electrode at its back connected to a current amplifier for signal detection. A laser and detector were placed 10 mm above the bottom electrode (red dot) to trigger the recording of the signal. The current signal was amplified and recorded by data acquisition (DAQ) system. (b) The recorded signal from a 50 µL water droplet sliding on OTS coated glass. The first peak corresponds to the accumulated charge in a droplet while the second (inverse peak) represents the deposited charge due to slide electrification in addition to charge on a drop. The red shaded area was integrated to determine the accumulated charge on the droplet. (c) Advancing and receding contact angles of water drops on various hydrophobic coatings on glass. Error bar represents standard deviation of 3 measurements.*

To measure the charge (Q) deposited by the liquid/solid phase we used a tilted plate setup enclosed in a temperature controlled chamber (Fig. 1a, Supplementary Video S1 & S2).[12] Drops/ice slide down a hydrophobized glass plate. A copper electrode of 20 mm length at the back of the substrate was connected to a current amplifier to detect the drop charge. It was positioned at a drop/ice sliding distance of 40 mm.[2] While a droplet slides down the solid surface it acquires an electrical charge and leaves an opposite charge on the surface. The sliding distance (40 mm) was chosen such that drop/ice is saturated with charge.[44] We measured the resulting capacitive current induced when the charged drop slides over the electrode. When a charged drop passes over the electrode, it generates a bipolar induced current signal (Fig. 1b) due to the generation of image charges. For a water droplet sliding over octyltrichlorosilane (OTS) coated glass, the first peak is a positive current peak due to the accumulated positive droplet charge. Electrons are flowing into the electrode. After the drop passes the center of the electrode, the current reverses to a negative peak due to electrons flowing back to ground. We integrate the first half of the bipolar current peak to estimate the amount of accumulated charge (red shaded area in Fig. 1b). Because the contact area of sliding ice cannot be determined reliably, owing to possible microscopic roughness and the presence of a quasi-liquid layer, we report total charge rather than charge density. The surface is initially neutralized with an ion gun thus the recorded drop charges correspond to the charges of the first drop sliding on a previously uncharged surface. Advancing and receding contact angles of water drops on various hydrophobic coating are given in Fig. 1c. Characterization and properties of surfaces and liquids used are given in Supplementary Fig. S1 & S2 and Table 1.

As an example, we present the drop charges measured using OTS-coated glass, as these surfaces lead to strong charge separation (Fig. 2). Within a temperature range of 0 to 20°C, a 50 $\mu L$ water drop accumulates a charge of around 1 nC. Following the common theory, we attribute this signal to adsorbed hydroxide ions ($OH^-$).[1,2,21,45,46] Water contains hydroxide ions and hydronium ions ($H_3O^+$) due to self-ionization. During its interaction with the hydrophobic surface, $OH^-$ ions are adsorbed at the surface-liquid interface and form the electrical double layer (EDL).[23] Upon dewetting, some $OH^-$ ions stay on the surface, resulting in a negatively charged surface. The water droplets accumulate $H_3O^+$ ions, resulting in net positive charge. We would like to point out that it has not yet been proven that hydroxide ions cause slide electrification, and that other ions (e.g. carbonate) may also be involved.[47]

*Table 1: Properties of liquids used in the study at 20°C.*

| Property | Water | Formamide | Diiodomethane | 1-Bromonaphthalene |
|---|---|---|---|---|
| **Surface tension (mN/m)** | 72 | 58 | 51 | 44 |
| **Dielectric constant ($\epsilon_r$)** | 80 | 111 | 5.3 | 4.8 |
| **Melting point (°C)** | 0 | 2 | 6 | −2 |
| **Density (g/cm³)** | 0.997 | 1.13 | 3.306 | 1.48 |
| **Viscosity (mPa·s)** | 0.89 | 3.3 | 2.6 | 4 |

Ice showed lower and a constant charge ($0.5 - 0.8$ nC) at temperatures $\leq -5$°C. Near the melting point, the charge (Fig. 2) was even higher with ice (2.2 nC) than with a water droplet ($0.8 - 1.2$ nC). The pronounced increase in deposition rates for ice near 0° C indicates that in addition to ionic transfer of surface charges additional processes contribute in this regime.[35] Similar charging behaviour – high charging for the liquid, low charging for ice at or below $-5$° C and a peak near the melting point – was observed for perfluorooctyltrichlorosilane (PFOTS) and chloro(dimethyl-tridecafluorooctyl)silane (PFOMS) coated substrates (Fig. 3). This difference did not arise because of different sliding velocity as velocity remained almost constant across phase and temperature (Supplementary Fig. S3).

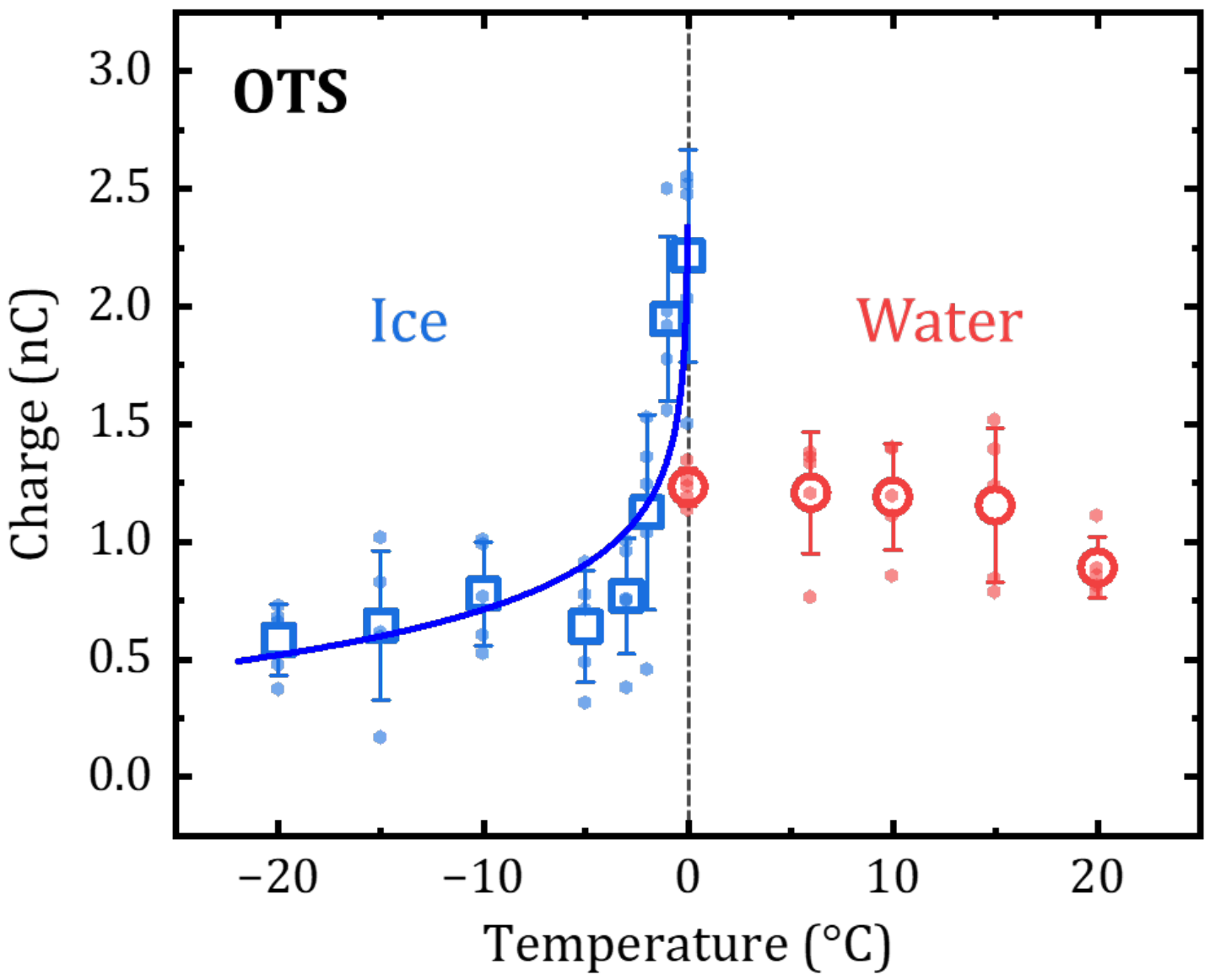

*Fig. 2: Charge of water drops (red) and ice (blue) created on OTS coated glass upon varying the temperature (±1°C). The width of the ice and water droplet was around 5 mm. Blue curve represents the weighted fitting of the logarithmic form $Q = A \ln(T_m - T) + B$. Fitting parameters and reduced $\chi^2$ values are given in Supplementary Table S2.*

The two charging regimes for ice and water and the peak in between indicate that two different charge transfer processes prevail. For water at T ≥ 0°C the results agree with the deposition of anions from the electric double layer. Below 0°C, for ice, the formation of a diffuse layer is hindered. It is established that a liquid-like layer exists at the ice–vapor and at ice-solid interfaces.[48] This quasi-liquid layer, which forms due to surface premelting,[35-37,49,50] can enable $OH^-$ ion transfer and deposition onto the surface.[51] The thickness of this layer is highly temperature dependent.[52] Below approximately −16°C, it has been reported to decrease to the order of a bilayer of water molecules.[52] This layer is much thinner than the typical Debye length of an aqueous electric double layer (∼100 nm). Moreover, our measurements show no temperature dependence of charge deposition that would correlate with changes in liquid-like layer thickness nor residual temperature gradients in the ice. However, protons and hydroxides still exist which can diffuse and cause significant charging. In addition, formamide showed a charging behaviour similar to that of water (Supplementary Fig. S4).

A possible second mechanism, which may dominate below 0°C involves electron ($e^-$) transfer driven by direct contact between the ice and the substrate during sliding.[53] In this scenario, the overlap of electron clouds at the interface facilitates net electron transfer from the ice to the surface, leading to negative surface charging. Near 0° C, the simultaneous presence of both solid and liquid water phases can promote a hybrid charging mechanism, where electron transfer coexists with ionic ($OH^-$) transfer, enhancing the overall deposition rate and thus, increasing the charge on ice.[35]

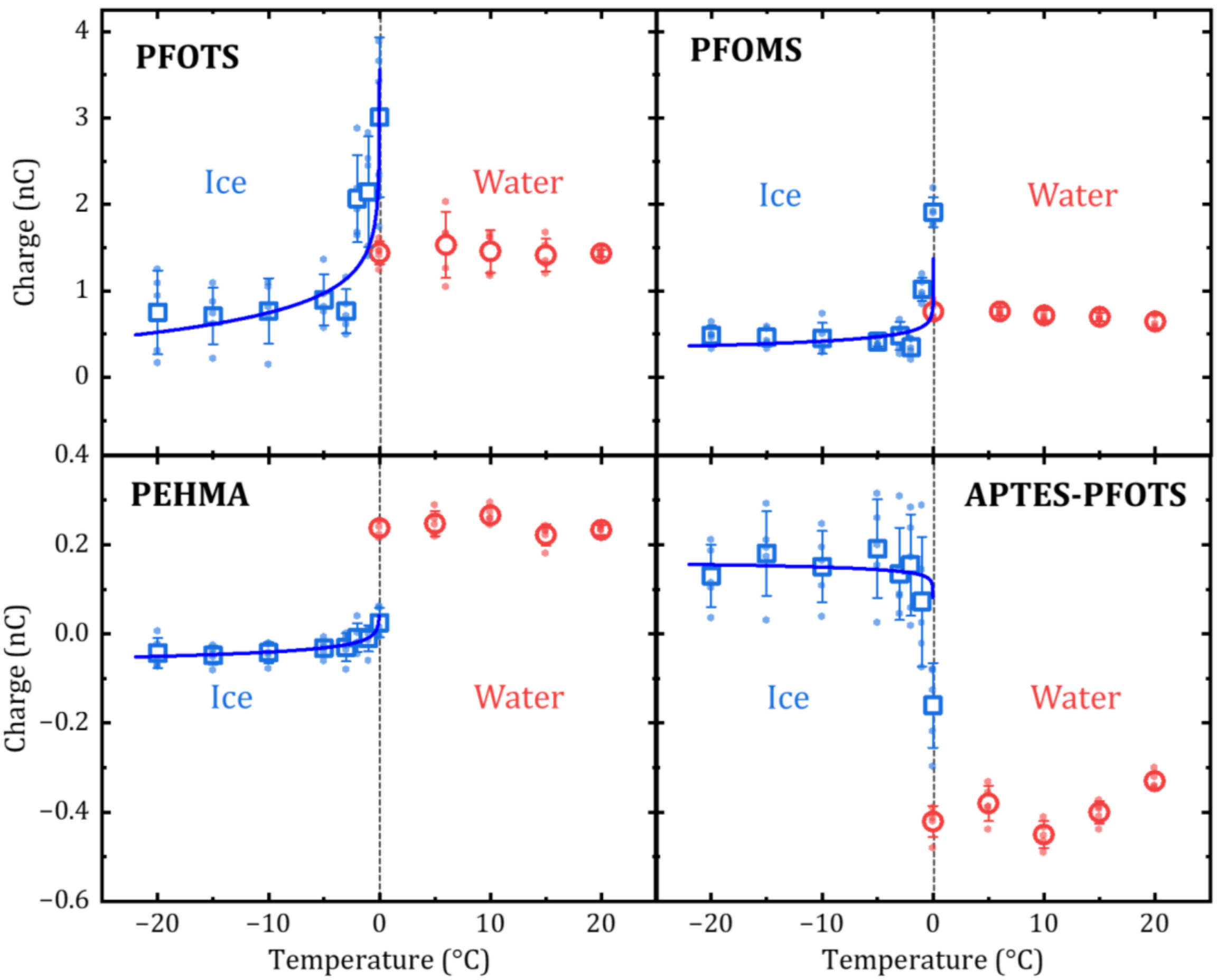

*Fig. 3: Drop charge versus temperature* (±1℃) *for water/ice measured on a PFOTS, PFOMS, PEHMA and APTES-PFOTS coated glass. Blue curves again represent the weighted fitting of the logarithmic form* $Q = A\ ln(T_m - T) + B$.

Ice and water do not always charge in the same polarity (Fig. 3). For poly(2-ethylhexyl methacrylate) (PEHMA)-coated glass, interaction with a water droplet results in a positive charge of +0.22 to + 0.27 nC, which is consistent with $OH^-$ adsorption from the liquid onto the surface (Fig. 3). In contrast, when the same coating interacts with ice at temperatures at or below −5° C, a net negative charge of −0.03 to − 0.05 nC is accumulated (Fig. 3). This polarity switch cannot be explained by ions alone, pointing towards an alternate mechanism. The effect is reversed for (3-aminopropyl)triethoxysilane (APTES)–PFOTS coatings, where water charges negatively and ice positively (Fig. 3). The amine group gets protonated from the sliding water drop (R-$NH_2$ + $H_3O^+$ → R-$NH_3^+$ + $H_2O$), resulting in a net positive surface charge when contacted by water at near-neutral pH; the surface $pK_a$ of primary APTES layers is ~ 9.6 so a significant fraction is protonated under our conditions.[19,54] As a result, the drop charges negatively. However, ice may transfer $e^-$ to the surface, making it negatively charged. At 0°C, the measured charge for both coatings lies between that of water and ice. This is consistent with a regime where ion and electron transfer occur simultaneously but in opposite directions, partially cancelling one another and yielding a reduced net charge.

The polarity inversion suggests that ion transfer from the quasi-liquid layer does not fully account for ice charging below −5°C. While direct evidence of $e^-$ transfer in such systems remains lacking, earlier work has shown that ice–metal contact electrification depends on the Volta potential (V) difference.[55-58] Reported logarithmic dependences of the form $V \propto \ln(T_m - T)$, where $T_m$ is the melting temperature, suggest a change in charging as the melting point is approached.[57] To assess compatibility with this logarithmic form, we performed weighted fits of the function $Q = A \ln(T_m - T) + B$ to the data, where both A and B have units of Coulombs (Blue curves in Fig. 2 & 3). As shown in Supplementary Table S2, the reduced $\chi^2$ values vary across substrates ($\chi_r^2 \approx 1.1$ for OTS and PFOTS, $\chi_r^2 = 3.26$ for PFOMS, and $\chi_r^2 \ll 1$ for PEHMA and APTES-PFOTS). Additionally, coefficients are also weakly constrained. Thus, the logarithmic form cannot be rejected, but neither can it be uniquely confirmed as the physically required dependence, particularly given the non-uniform behavior across surfaces.

Moreover, the polarity reversal between PEHMA and APTES-PFOTS surfaces constitutes a violation of the transitive property of the triboelectric series: if charging were governed by a single ordered series, the polarity of deposition should follow a consistent ranking across all surfaces and liquids. The observed reversal therefore is difficult to reconcile with a single surface-potential-based mechanism. We note that surface and liquid contamination can influence charge polarity.[59]

However, the systematic and reproducible nature of the polarity patterns across surfaces makes contamination an unlikely sole explanation for the observed behavior.

On their own, phase-dependent charging in polar liquids do not confirm distinct mechanisms; however, charging behavior in a non-ionizing liquid does provide an important clue. We carried out charge measurements with nonpolar liquids which still have a high surface tension, such as diiodomethane and 1-bromonaphthalene. In contrast to the polar liquids tested, the number of free ions and the formation of an EDL should be negligible.[43] Moreover, our tested polar liquids are self-ionizing in nature while non-polar liquids are not. Drop charging was reduced to 30-50% compared to that of water (Fig. 4). For the non-polar liquids, drop charging decreased to ≤25%, presumably because the formation of surface charges and an electric double layer is suppressed.[43] Even charge reversal was observed for OTS and PEHMA coated surfaces. Significant charge with nonpolar liquids suggests a presence of an alternate mechanism (Fig. 4). Because nonpolar liquids have a relatively low dielectric constant ($\epsilon_r$ = 5.3 for diiodomethane and 4.8 for 1-bromonaphthalene) as compared to polar liquids ($\epsilon_r$ = 80 for water and 111 for formamide), the electrostatic energy cost of separating charges is high ($U \sim 1/\epsilon_r$). The much larger $U$ in nonpolar liquids implies that ion pairs are much more tightly bound and spontaneous ion separation is strongly suppressed.[43] Thus, higher energy cost suppresses ion-based charge separation during sliding. However, other charging pathways (i.e., electrons) can still operate and result in significant deposition.

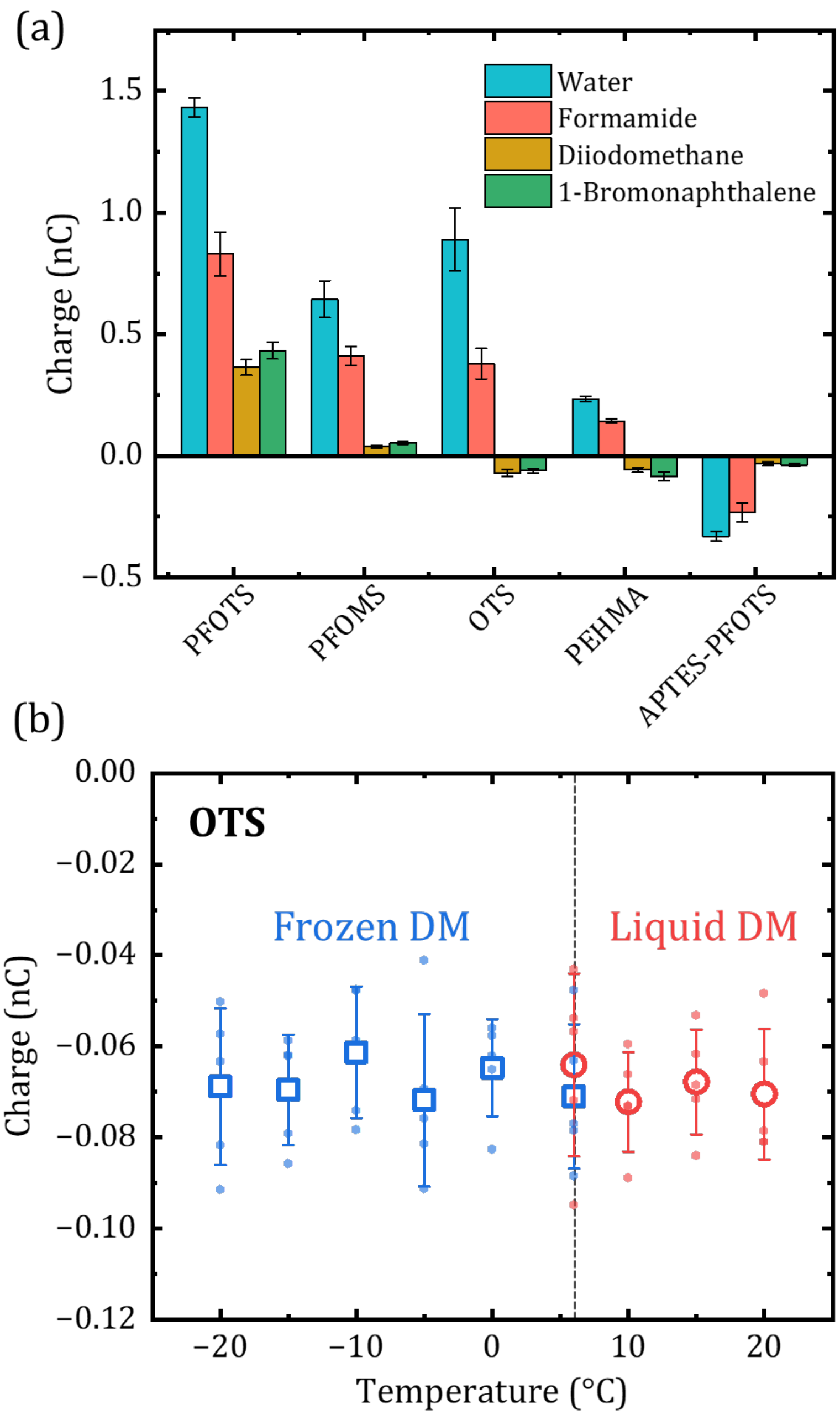

*Fig. 4: (a) A comparison of charge on polar (water and formamide) and nonpolar (diiodomethane and 1-bromonaphthalene) drops at 20°C on various surfaces. (b) Charge on liquid and frozen phase diiodomethane on OTS coated glass showed consistent behavior without any peak at melting point (6°C, black dotted line).*

In Fig. 4a, we compare drop charges measured with different liquids on surfaces with different chemical functionalities. On PFOTS, PFOMS and APTES-PFOTS coatings, all liquids consistently deposited charges of the same polarity, although the magnitudes differ. Conversely, for OTS and PEHMA coatings, a polarity reversal was observed when switching from polar to nonpolar liquid (Fig. 4a). These results demonstrate that hydroxyl groups (for PFOTS and PFOMS) or protons (APTES-PFOTS) are not the only charges, which are transferred, because they cannot explain the drop charges in diiodomethane and bromonaphthalene.

Furthermore, if water contamination was responsible for the charging by diiodomethane and bromonaphthalene, the polarity would remain consistent with water droplets across all surfaces. We further performed Karl Fischer Coulometry to quantify residual water and found <0.05% water in the nonpolar liquids and <0.4% in formamide. The reversal therefore supports an additional mechanism at play in nonpolar liquids.

We further compare charge deposition with a nonpolar liquid and its frozen phase (Fig. 4b). A liquid diiodomethane droplet (50 $\mu L$) at room temperature accumulated a charge of around $-0.07$ nC on OTS coated glass. It was independent of temperature for the tested range of -20°C to 20°C. No significant change was detected at the melting point of 6°C between liquid and frozen diiodomethane, which is different from our observation with polar liquids (Fig. 2a). Similar charging behavior was also observed for other hydrophobic substrates and nonpolar liquids (Supplementary Fig. S5 & S6). The fact that a discontinuity is observed for polar liquids but not for the nonpolar liquids suggests that it is not a trivial consequence of a phase transition. It rather reflects a fundamental difference between charging mechanism. We note, however, that temperature-dependent Volta-potential data analogous to those reported for ice are not available for diiodomethane or 1-bromonaphthalene. Therefore, the present nonpolar-liquid data cannot be used as a direct test of the logarithmic Volta-potential dependence proposed in Ref.[57]. Nevertheless, the finite charging of nonpolar liquids, together with their weak phase dependence and occasional polarity reversal relative to water, supports the presence of an additional charge-transfer pathway beyond conventional ion transfer. It is consistent with the electron transfer hypothesis: as ions most likely do not contribute to charge deposition in nonpolar liquids, electrons may provide the dominant transfer pathway. Additionally, the charging magnitude of polar liquids consistently exceeds that of nonpolar liquids by approximately one order of magnitude across all five tested surfaces. This observation can naturally be explained by the contribution of ions present in polar, self-ionizing liquids but absent in nonpolar ones. While a purely electron-transfer mechanism cannot be strictly ruled out based on the absence of a melting-point discontinuity alone as the Volta potentials of nonpolar compounds may vary only weakly across melting.

To further analyze a possible contribution of electron transfer to charging, we plotted the average electronegativity of the liquids and the surface functional groups with the polarity of charge deposition. For the nonpolar liquids we found a correlation between the average electronegativity of liquid and the surface functional groups with the polarity of charge deposition. Specifically, the direction of charge transfer is correlated with the difference in electronegativity between the surface coating and the liquid molecules (Fig. 5). Here, the electronegativity difference is defined as $\Delta_{en} = \chi_s - \chi_l$, where $\chi_s$ and $\chi_l$ are the average Pauling's electronegativities for surface group and liquid molecule, respectively. $\chi_s$ and $\chi_l$ were calculated by taking arithmetic average of individual atoms' electronegativity.[60,61] Values are given in Supplementary Table S1. When the average electronegativity of the surface exceeds that of the liquid ($\Delta_{en} > 0$), electrons preferentially transfer from the sliding liquid to the surface, resulting in a net positive charge of the drop. Conversely, when the liquid has a higher electronegativity ($\Delta_{en} < 0$), electrons transfer from the surface to the liquid, leaving the droplet negatively charged. This correlation provides a straightforward explanation for the polarity reversals observed in Fig. 4a. For example, PFOTS and PFOMS surfaces, which possess more electron-withdrawing groups than the tested nonpolar liquids, can charge liquid positively due to electron donation from the liquid. In contrast, coatings such as OTS and PEHMA, with lower effective electronegativity relative to the liquid, charges liquid negatively.

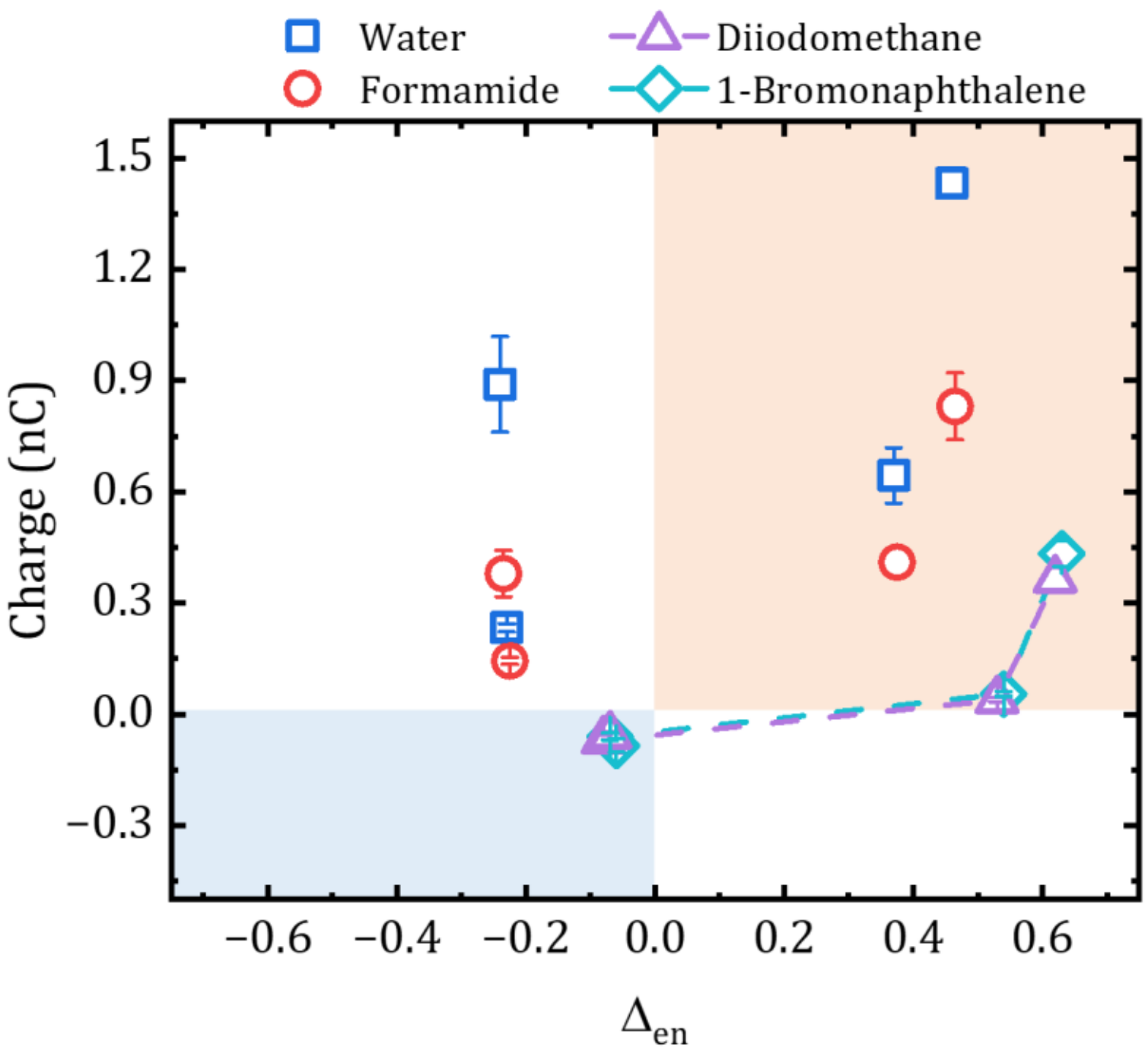

*Figure 5: liquid charge variation (20°C) variation with electronegativity difference ($\Delta_{en}$) for polar and nonpolar liquids. The red shaded area represents where droplet charge and $\Delta_{en}$ are positive while the blue area represents negative values.*

However, polar liquids do not follow this trend, likely due to the dominant contribution from ion transfer. Additionally, no direct correlation was observed between the amount of charge deposition and $\Delta_{en}$, as charge deposition is influenced by multiple factors, including surface wettability, droplet velocity, sample preparation, and surface roughness. Furthermore, the electronegativity of the outermost exposed surface groups is likely to play a more significant role in the charge transfer process than that of the inner functional groups. Additionally, water and non-polar liquids in their frozen states generally follow the electronegativity-difference trend, except for ice on OTS, which may indicate a still dominant contribution from a quasi-liquid interfacial layer (Supplementary Fig. S7). These results suggest that interfacial electron transfer is governed by relative electronegativity differences, and provide a predictive descriptor for charge polarity in liquid–solid triboelectrification. Electronegativity, as used here, serves as a molecular-level proxy for the same driving force that underlies Volta potential differences at the condensed-matter level. Where Volta potential data for the relevant liquid-solid pairs become available, they may provide a more rigorous quantitative test of this picture. While our observations strongly indicate a contribution from electron transfer as an additional mechanism, direct evidence for electron involvement remains lacking.

## Conclusion

Our study indicates that slide electrification is governed by a combination of two fundamental charge transfer mechanisms, with their relative contributions determined by the liquid's polarity, ionization, phase, temperature, and the electronegativity difference between liquid and surface. Polar liquids primarily transfer ions in their liquid state, whereas an additional mechanism dominates in frozen states and in nonpolar liquids. Near the melting point of polar liquids, both mechanisms can operate simultaneously, resulting in enhanced or intermediate charge deposition depending on the surface and liquid properties. These findings establish a unified framework for understanding charge generation in liquid–solid interactions and highlight the fundamental role of an additional mechanism alongside ions in governing surface charging across diverse materials and conditions.

## Methods

**Experimental details:**

The ice was frozen in a temperature-controlled chamber integrated within a glove box (Supplementary Fig. S8(a) & (b), Supplementary Video S1 & S2), which enabled experiments to be carried out without opening the chamber. Consequently, the freezing, transfer, and placement of the ice on the tilted plate were performed entirely inside chamber, thus avoiding exposure to ambient air and preventing thermal perturbations. The temperature of the chamber was maintained within ± 1$^{o}$C. Four different liquids were used: deionized water (Milli-Q, 18.2 MΩ.cm), formamide (99.5%, AppliChem GmbH), diiodomethane (>99%, Thermo Scientific) and 1-bromonaphthalene (>95%, Sigma Aldrich). The structures of liquids are given in Supplementary Fig. S8(c). A 50 $\mu L$ droplet was frozen on a hydrophobized silicon wafer to ensure a smooth, flat interface (Supplementary Fig. S8(d) & (e)). The frozen droplet was carefully removed from the silicon wafer with a metal tweezer. The same hydrophobic coating on glass was used for charge measurement to avoid cross-contamination. Any residual temperature mismatch after transfer is expected to equilibrate rapidly upon contact with the substrate, with equilibration over sub-millimeter to millimeter length scales within ~1 s. Within 1 s, the piece of ice equilibrates with the glass sample over a depth of $\Delta x \approx \sqrt{\alpha t}$ , where $\alpha$ is the thermal diffusivity. For water, $\alpha$=1.4×10$^{-7}$ m$^2$/s, for ice at 0°C $\alpha$=1.2×10$^{-6}$ m$^2$/s. Glass has a higher thermal diffusivity. After t=1 s, the temperature should have equilibrated due to heat flow over a distance of approximately 0.4 mm for water and 1 mm for ice. Humidity inside chamber at different temperatures is reported in Supplementary Fig. S9(a). Additionally, we did not observe any condensation or frosting even at sub-zero temperatures (Supplementary Fig. S9(b)). For charge measurements, liquid droplet/ice (50 $\mu L$) was dispensed on a hydrophobic glass surface on a tilted plate (50°). A copper electrode of 20 × 25 mm$^2$ sensing area and 3 mm thickness was placed on the backside of the surface at the sliding distance of 40 mm. The electrode was 20 mm wide and thus significantly larger than used in the literature.[12] The electrode was also larger than the drop size to avoid an overlapping of the bi-polar peaks. If the two peaks overlap, they partly cancel each other out, which leads to an underestimation of the charge. It is necessary to ensure two completely separated peaks for correct charge determination as shown in Fig. 1b. The remaining part of the glass plate was placed on a grounded metal plate. The electrode was connected to a current amplifier (DDPCA-300, FEMTO) and data acquisition system - DAQ (USB-6366, NI) to record the current signal. Amplification used was in the range of 10$^7$ to 10$^9$ V/A depending upon the sample with rise times of 0.8 ms to 2.3 ms. A light barrier formed by a laser and a detector (placed 10 mm above the electrode position) was employed to detect the incoming droplet/ice to trigger the recording of the current signal. All components of the setup were kept at the same temperature, except the amplifier and DAQ. All the surfaces were neutralized by Zerostat anti-static ion gun (Sigma Aldrich) before the experiments to neutralize the surface from any residual charge. Liquid droplets and ice were handled with grounded metal syringe and metal tweezer, respectively to avoid any pre-charging. Each experiment involved at least 5 measurements on different surfaces, which were utilized to calculate of the standard deviation. Additionally, surface was neutralized with ion gun before each measurement.

**Preparation of surfaces:**

Glass slides (76 × 26 × 1 $mm^3$, soda-lime, epredia) were used as a substrate for various hydrophobic coating. The protocol for preparation is available in literature.[12,19] Glass slides were first cleaned with ethanol and isopropyl alcohol under 10 min of ultra-sonication and blow-dried with nitrogen. $O_2$ plasma (Diener electronic, with 300 W and 0.3 bar for 10 minutes) was used to activate the surface before coating. Activated glass slides were placed inside a vacuum desiccator with 1 mL vial of 1H,1H,2H,2H-perfluorooctyltrichlorosilane (PFOTS, 97%, TCI) solution. The desiccator was evacuated with a vacuum pump to less than 50 mbar for 10 min – allowing PFOTS vapor to fill the entire chamber. Then the reaction was allowed to take place for 20 minutes at constant pressure after closing the chamber. The PFOTS-coated slides were rinsed with ethanol multiple times to rinse off unreacted silane[62] and dried with nitrogen before using them for charge measurements. The same process was followed for octyltrichlorosilane (OTS, 97%, Sigma-Aldrich) and chloro(dimethyltridecafluorooctyl)silane (PFOMS, >94%, TCI) coating. Additionally, we prepared mixed PFOTS-APTES-coated glass.[19] In this case, we started with a PFOTS primary coating with a reaction time of 10 min. Then (3-aminopropyl)triethoxysilane (APTES, 99%, Sigma-Aldrich) secondary coating was deposited in separate desiccator with 6 h of reaction time at 50 mbar chamber pressure.

We synthesized the grafted poly(2-ethylhexyl methacrylate) (PEHMA) surfaces in the same way as described previously.[63] Copper(I) bromide (>98%, TCI) was purified before the synthesis.[64] All other chemicals were used as delivered. 2-Ethyl methacrylate (EHMA,7.04g, 35.51 mmol, Sigma Aldrich), 4,4'-dinonyl-2,2'-dipyridyl (58.6 mg, 0.143 mmol, TCI), copper(II) bromide (6.6 mg, 0.029 mmol, Sigma Aldrich) and toluene (>99.8%, Fisher, Netherland)/DMSO (5.3 mL: 2.6 mL, puriss. p.a. dried <0.02%, Sigma Aldrich) were added to a flask, provided with a septum and degassed with argon for 1 h. After degassing, the copper(I) bromide was transferred with a syringe to the degassed (30 minutes) reaction reactors, containing the Br-SAM grafted substrate. The syringe was degassed with argon three times before using. The polymerization was performed at 30°C. After the desired reaction time, the reaction was terminated by rinsing the polymer brush surfaces two times with toluene, ethanol and DI-water once the desired thickness is reached (12 nm).

**Characterization:**

Contact angle measurements were carried out using a Krüss goniometer to evaluate surface wettability. Advancing and receding contact angles were determined by gradually adding and withdrawing liquid (5 µL/min) into a droplet of deionized water (5–35 µL) on the sample surface for 3 cycles (Fig. 1c). Surface topography and roughness were characterized using Atomic Force Microscopy in tapping mode (AFM, Dimension ICON FastScan). Supplementary Fig. S1 represents AFM topography and roughness characterization of all the samples.

The X-ray diffraction (XRD, Rigaku SmartLab, 1.54 Å wavelength) measurements for the all the frozen samples are given in Supplementary Fig. S2. All the liquids were frozen first on OTS-coated silicon wafers and subsequently transferred to the XRD chamber in liquid nitrogen (Formamide was frozen directly by liquid nitrogen). During XRD, temperature was kept constant at -20° C for water, diiodomethane and 1-bromonaphthalene while -50° C for formamide. The ice phase of water showed mostly hexagonal structure (Supplementary Fig. S2).[65]

The Karl Fischer Coulometry was performed with a Mettler Toledo C20 compact coulometer using a 1 mL syringe. 1 mL of the solvent was taken with the syringe. The analyte was weighted and the water content was measured by the Karl Fischer Coulometry. Every solvent was measured at least three times. Water content by volume of solvent found was 0.051 ± 0.001 % in diiodomethane, 0.024 ± 0.003 % in 1-bromonaphthalene and 0.38 ± 0.04 % in formamide.

## Author Contributions

H.J.B. and R.L. conceived the project. R.L. designed experiments and carried out charge measurements. B.L. and R.L. prepared and characterized samples. A.D.R. devised the measurement method. H.J.B. and W.S. supervised the project. R.L. wrote the manuscript. H.J.B., R.L., W.S., B.L. and A.D.R. discussed results, reviewed and revised the manuscript.

## Acknowledgement

This work was supported by the European Union's ERC Advanced Grant No. 883631 "DynaMo". Authors acknowledge Rüdiger Berger for providing the temperature-controlled chamber and Pravash Bista for help in initial setup.

## Competing interests

The authors declare no competing interests.

## Supplementary information

**Supplementary Information**

Supplementary Figs. S1–S10 and Table S1.

**Supplementary Video S1**

Measurement with water droplet on PFOTS coated glass at $50^{\circ}$ tilt angle and room temperature.

**Supplementary Video S2**

Measurement with Ice on PFOTS coated glass at $50^{\circ}$ tilt angle and $-5^{\circ}$C temperature.

# Mechanisms in Slide Electrification of Liquid and Frozen Drops on Hydrophobic Surfaces

## Supplementary Information

Rutvik Lathia[1,*]; Benjamin Leibauer[1]; Aaron D. Ratschow[1]; Werner Steffen[1] and Hans-Jürgen Butt[1,*]

[1] Max Planck Institute for Polymer Research, Ackermannweg 10, 55128 Mainz, Germany

*Corresponding authors: lathiar@mpip-mainz.mpg.de; butt@mpip-mainz.mpg.de

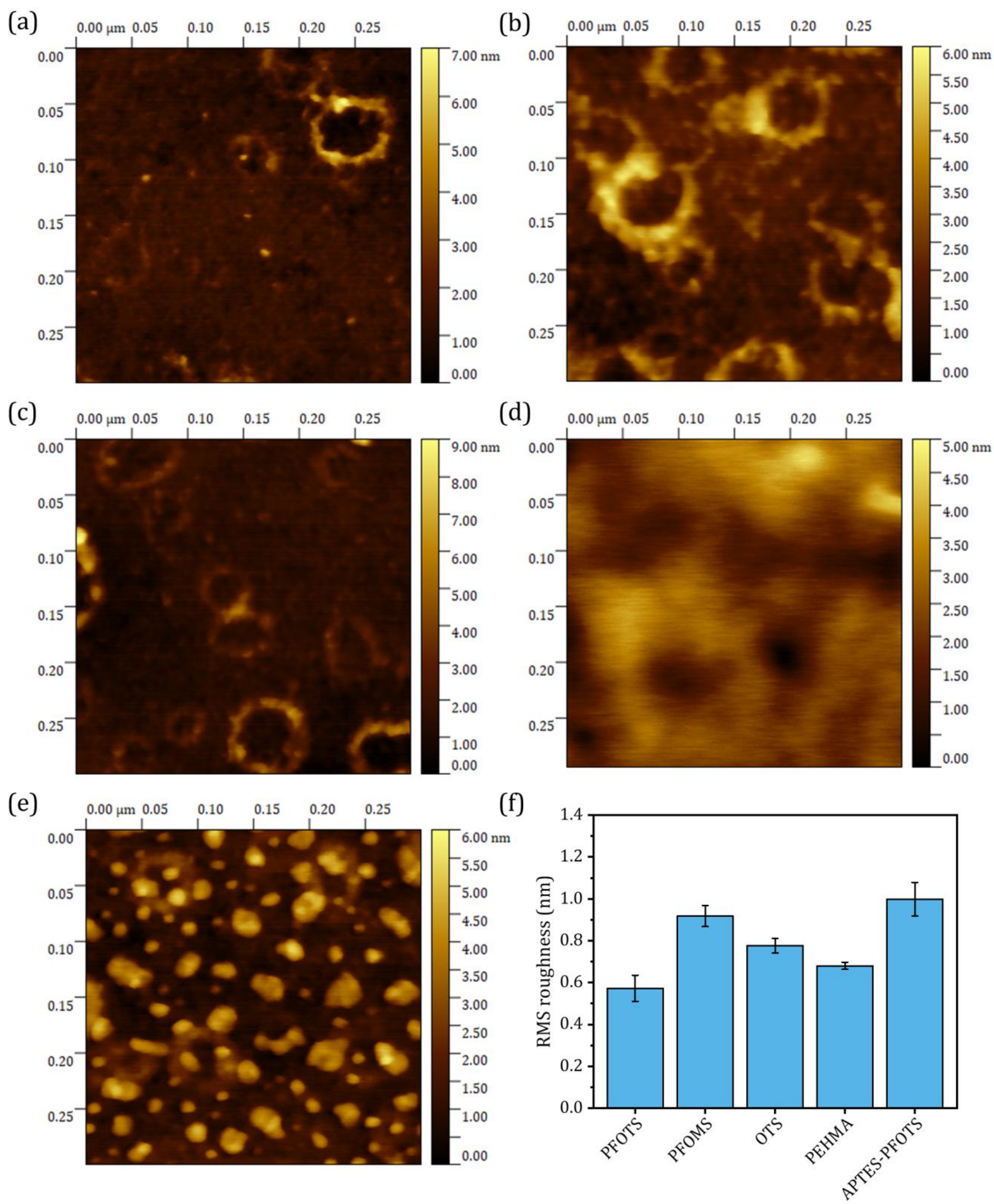

Supplementary Fig. S1: AFM topography images of (a) PFOTS (b) PFOMS (c) OTS (d) PEHMA and (e) APTES-PFOTS on glass. (f) RMS roughness for all the hydrophobic coatings. Error bar represents standard deviation of 3 measurements.

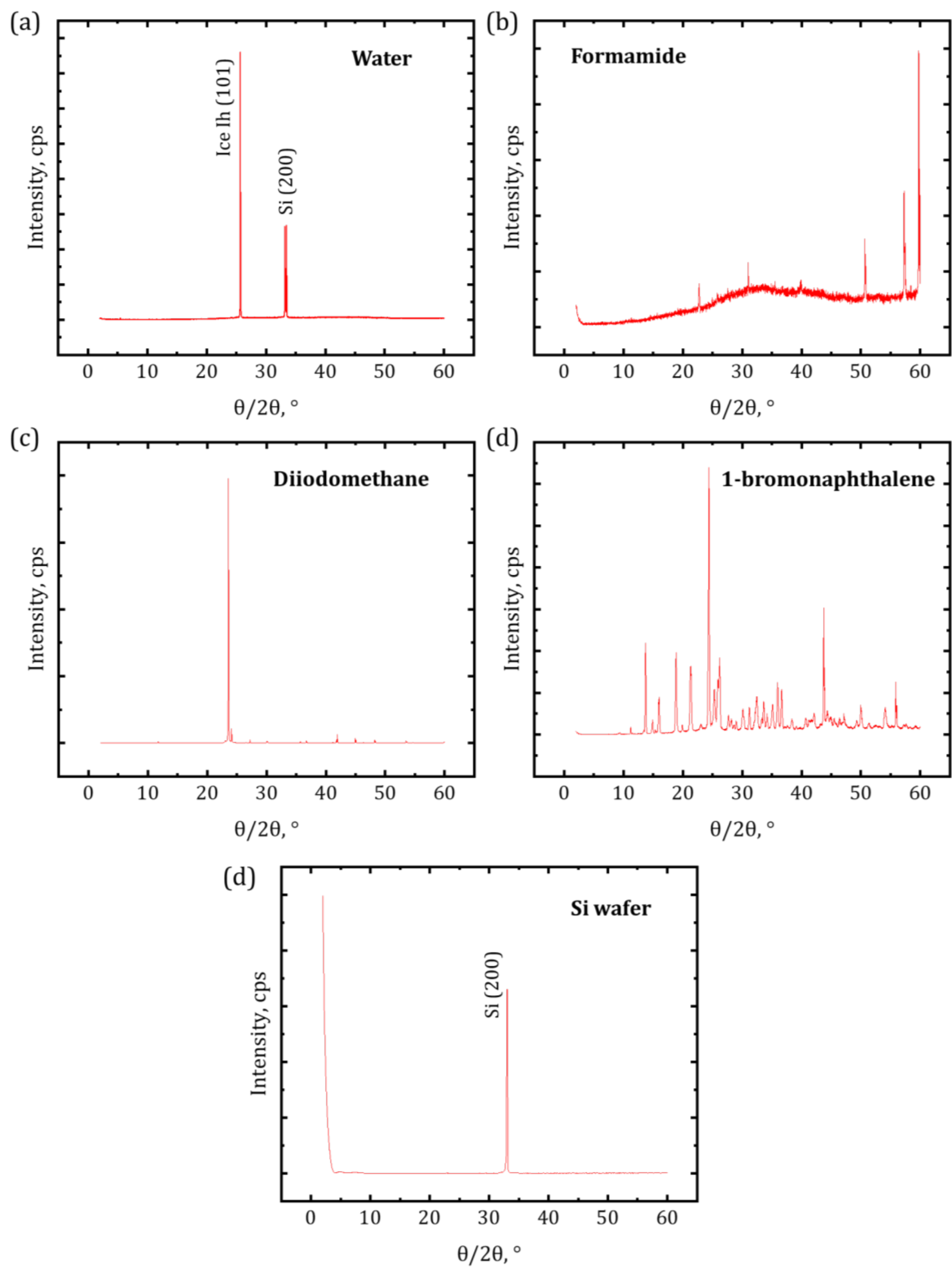

Supplementary Fig. S2: XRD diffractograms for all the frozen chemicals and OTS-coated Si-wafer. Formamide spectra were recorded at -50 °C while other all data were acquired at -20 °C.

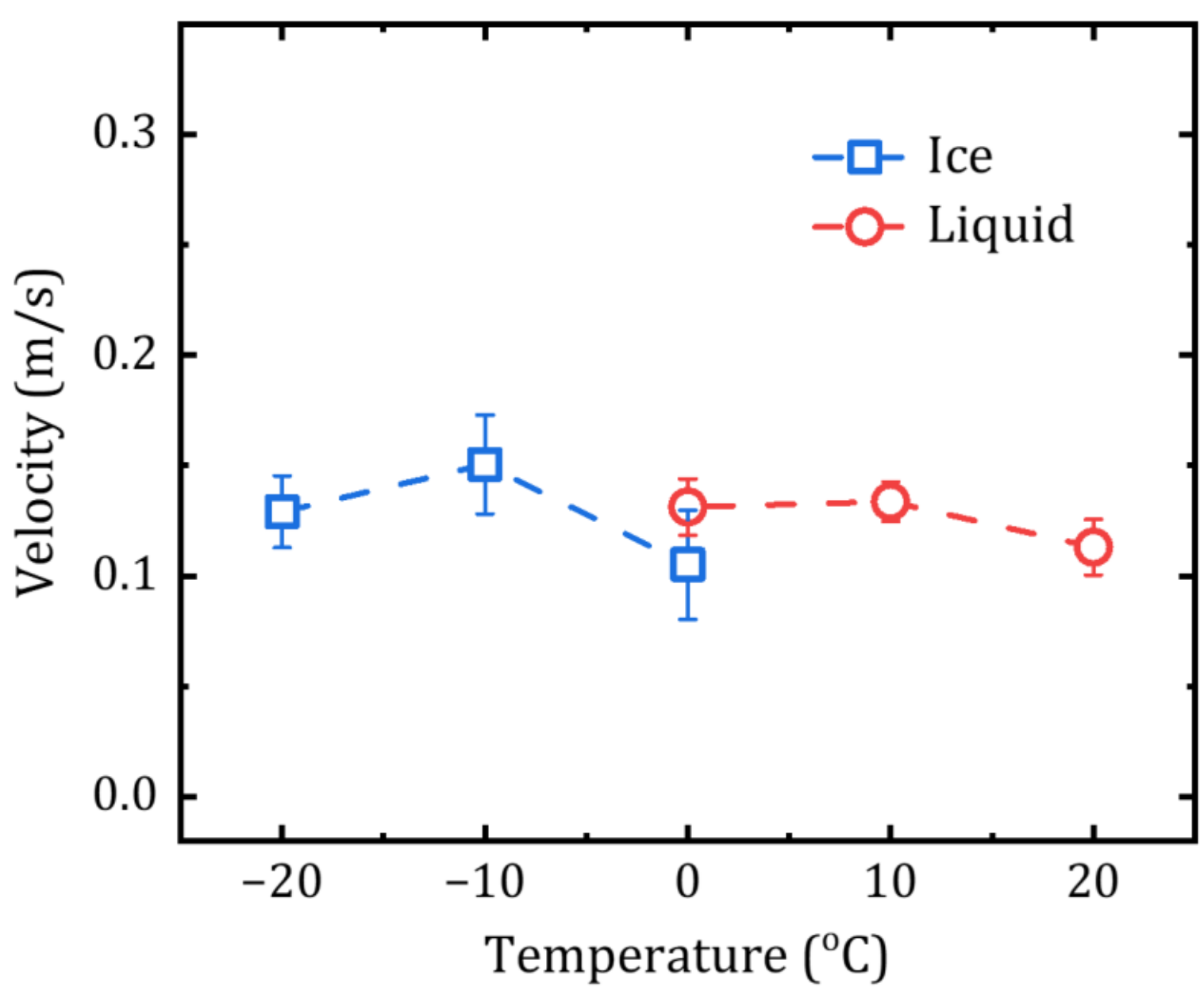

Supplementary Fig. S3: Velocity of water droplet of 50 μL volume and ice sliding over PFOMS coated glass.

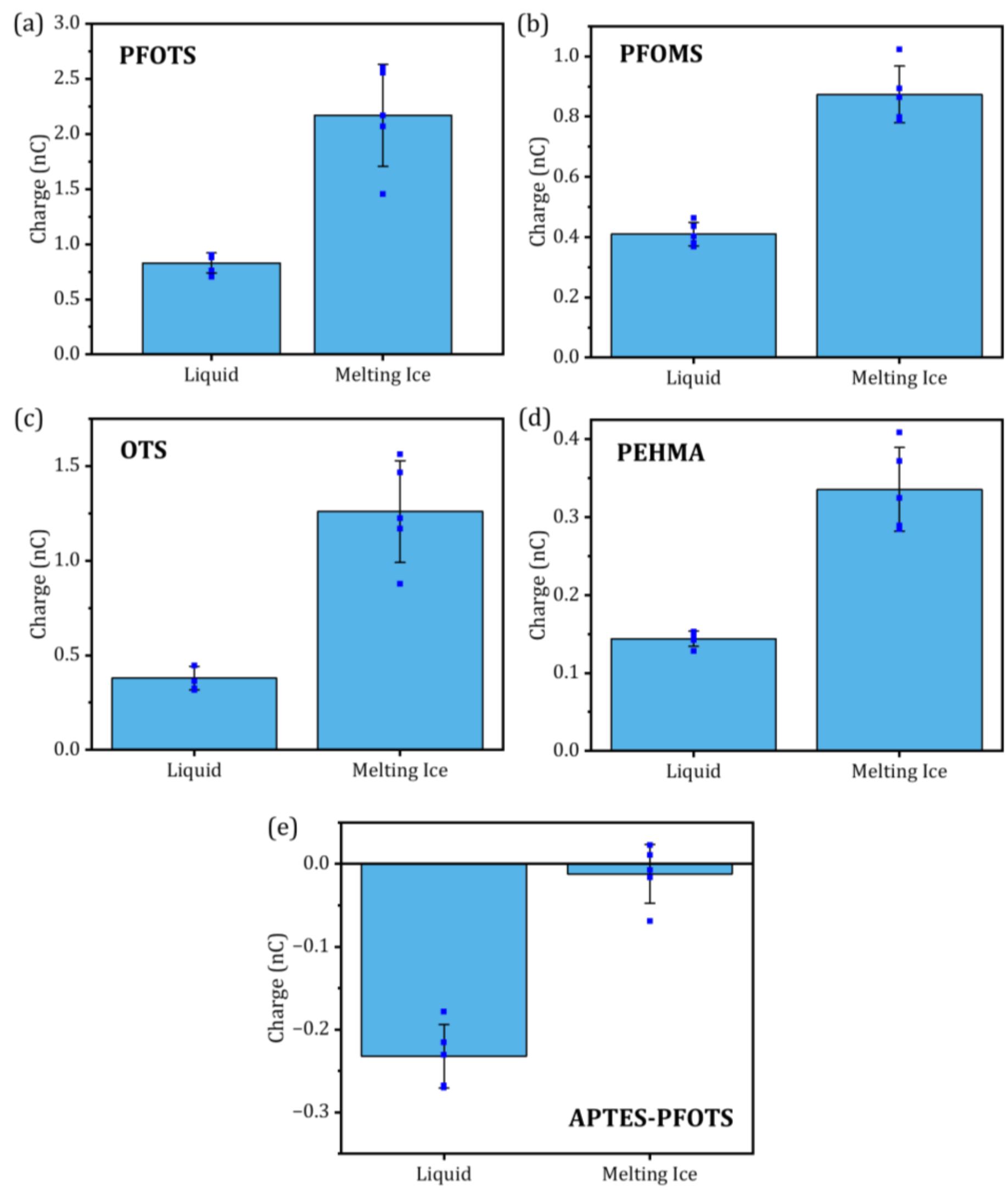

Supplementary Fig. S4: Drop charge of liquid (20°C) and melting formamide (at -20°C) measured on the five different surfaces. Formamide was frozen first with liquid nitrogen and then charging measurements were carried out as it was not freezing in a droplet form at -20°C. Thus, it was melting at -20°C.

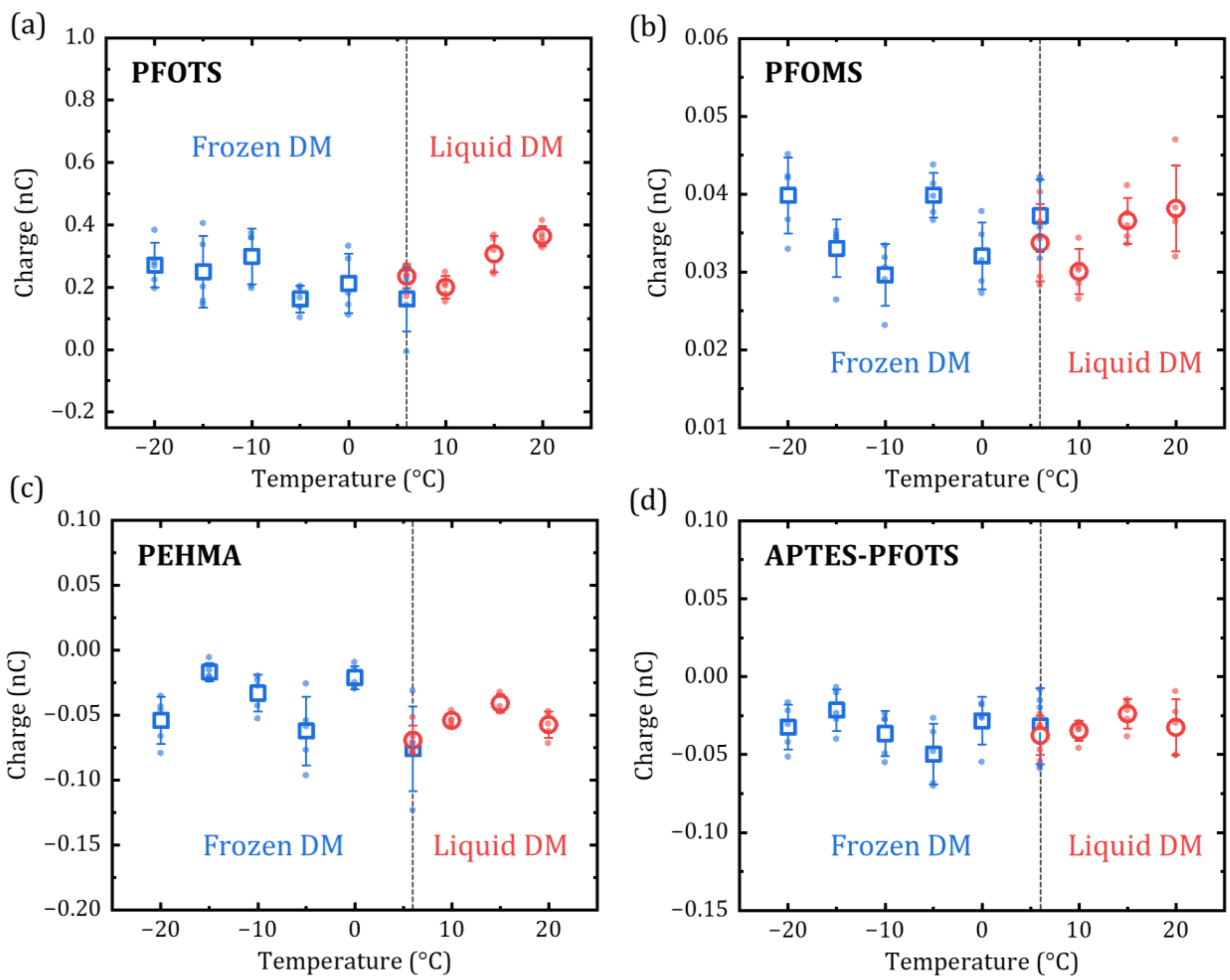

Supplementary Fig. S5: Drop charge versus temperature (±1°C) of liquid (red) and frozen (blue) diiodomethane on various surfaces.

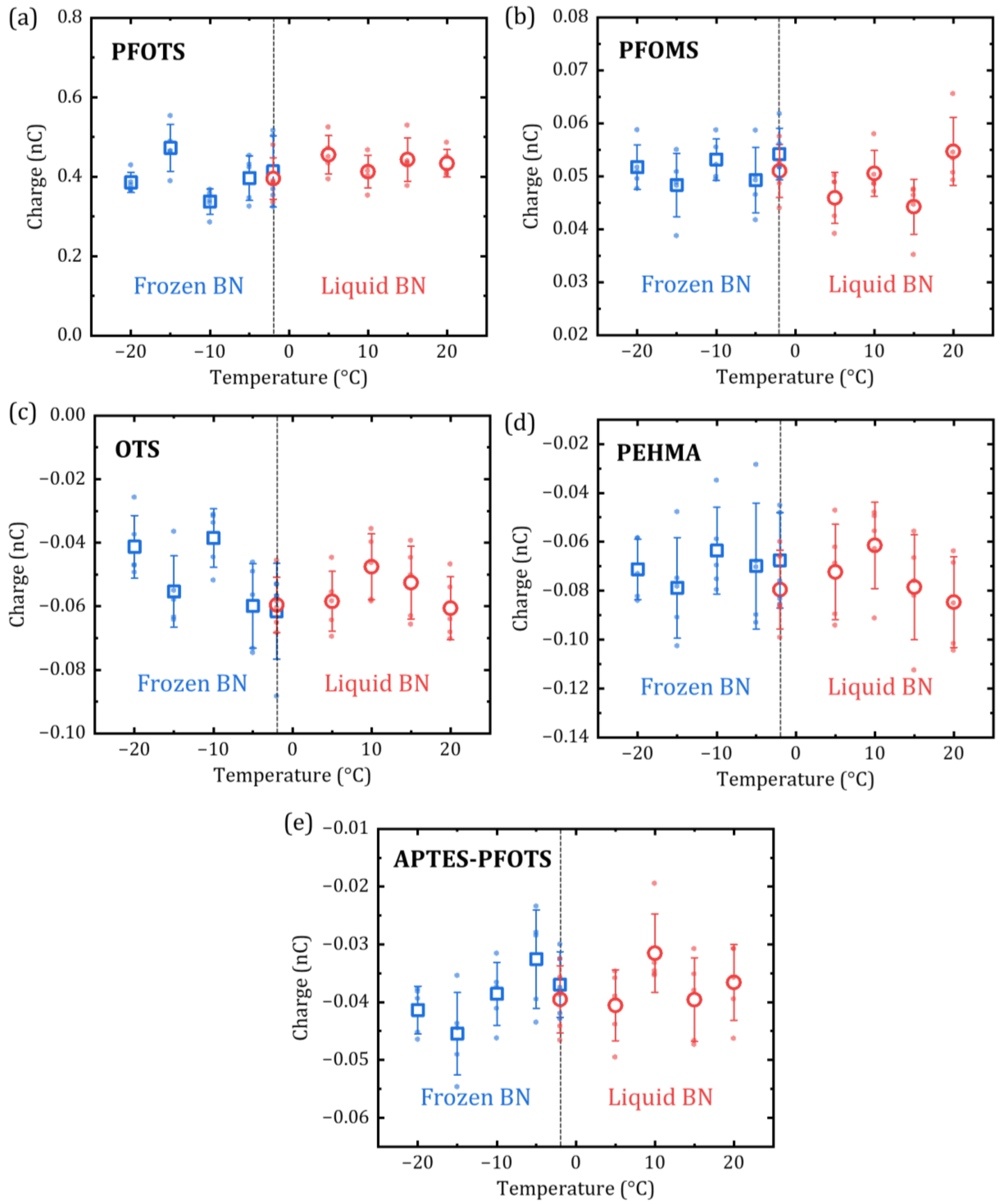

Supplementary Fig. S6: Drop charge versus temperature (±1°C) of liquid (red) and frozen (blue) 1-bromonaphthalene on various surfaces.

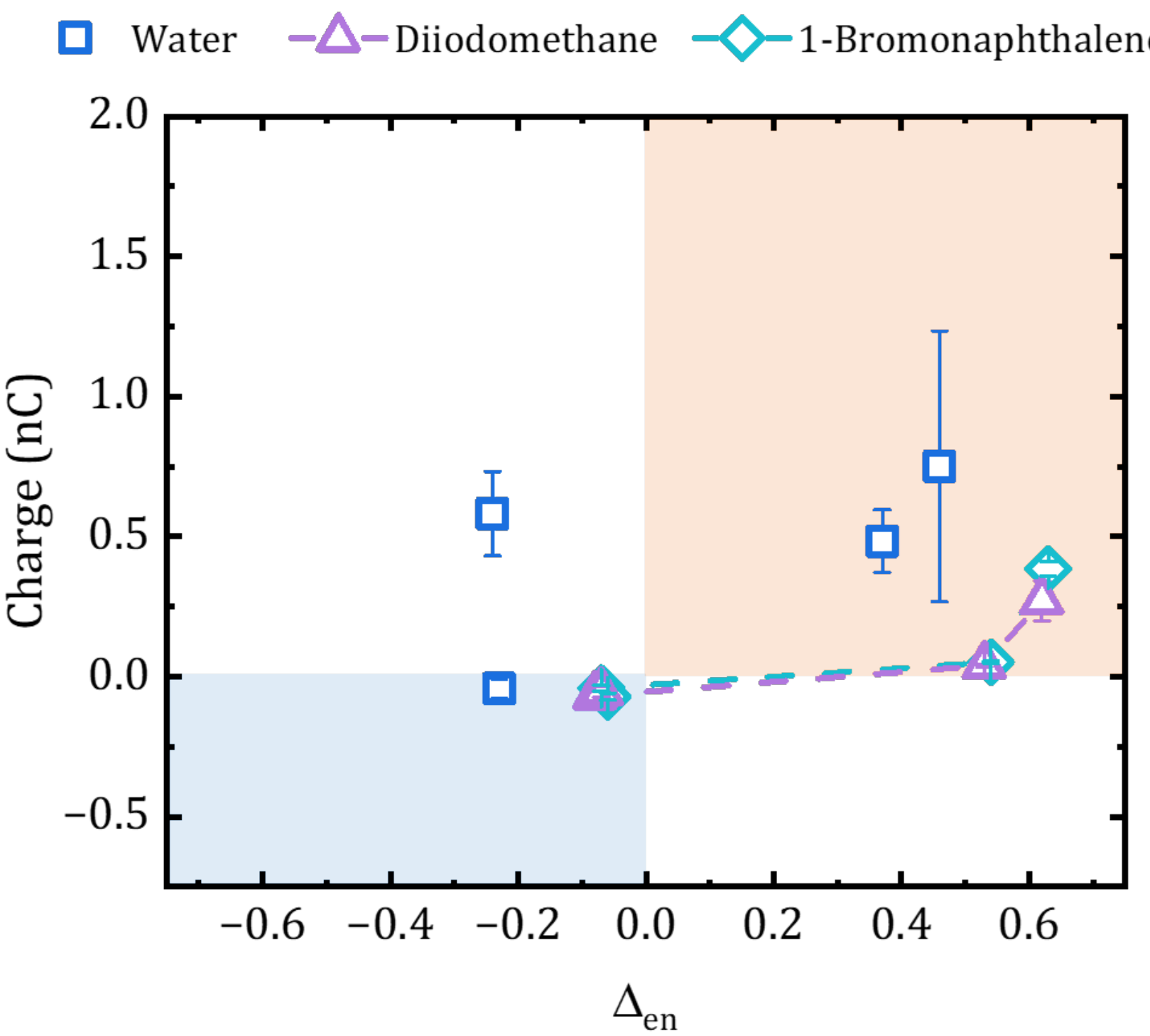

Supplementary Fig. S7: Frozen liquid charge variation (-20°C) variation with electronegativity difference ($\Delta_{en}$) for polar (Water) and nonpolar liquids. Formamide data points is not present as Formamide immediately started melting even at -20°C. The red shaded area represents where droplet charge and $\Delta_{en}$ are positive while the blue area represents negative values.

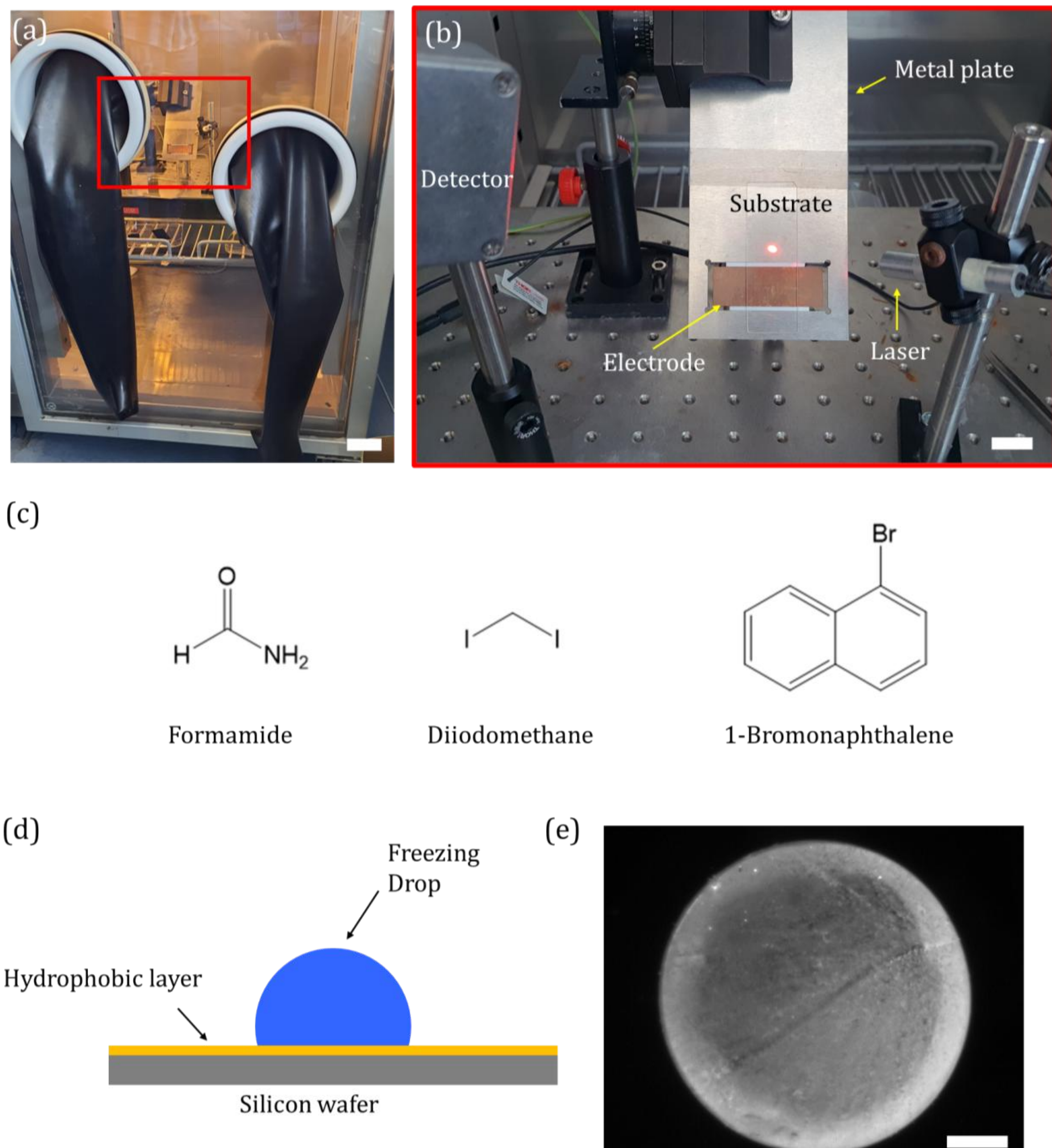

Supplementary Fig. S8: (a) Photograph of a temperature-controlled glovebox chamber (scale = 5 cm) and (b) charge measurement setup (scale = 2 cm). The freezing, transfer, and placement of the ice on the tilted plate were performed entirely inside chamber, thus avoiding exposure to ambient air and preventing thermal perturbations. (c) Chemical structure of the liquids used for characterizing electrification. (d) Droplets deposited on a hydrophobic layer–coated silicon wafer and subsequently frozen. (e) Inverted light microscopy view of the resulting frozen ice (50 $\mu$L) on PFOTS coated silicon wafer. Scale = 1 mm.

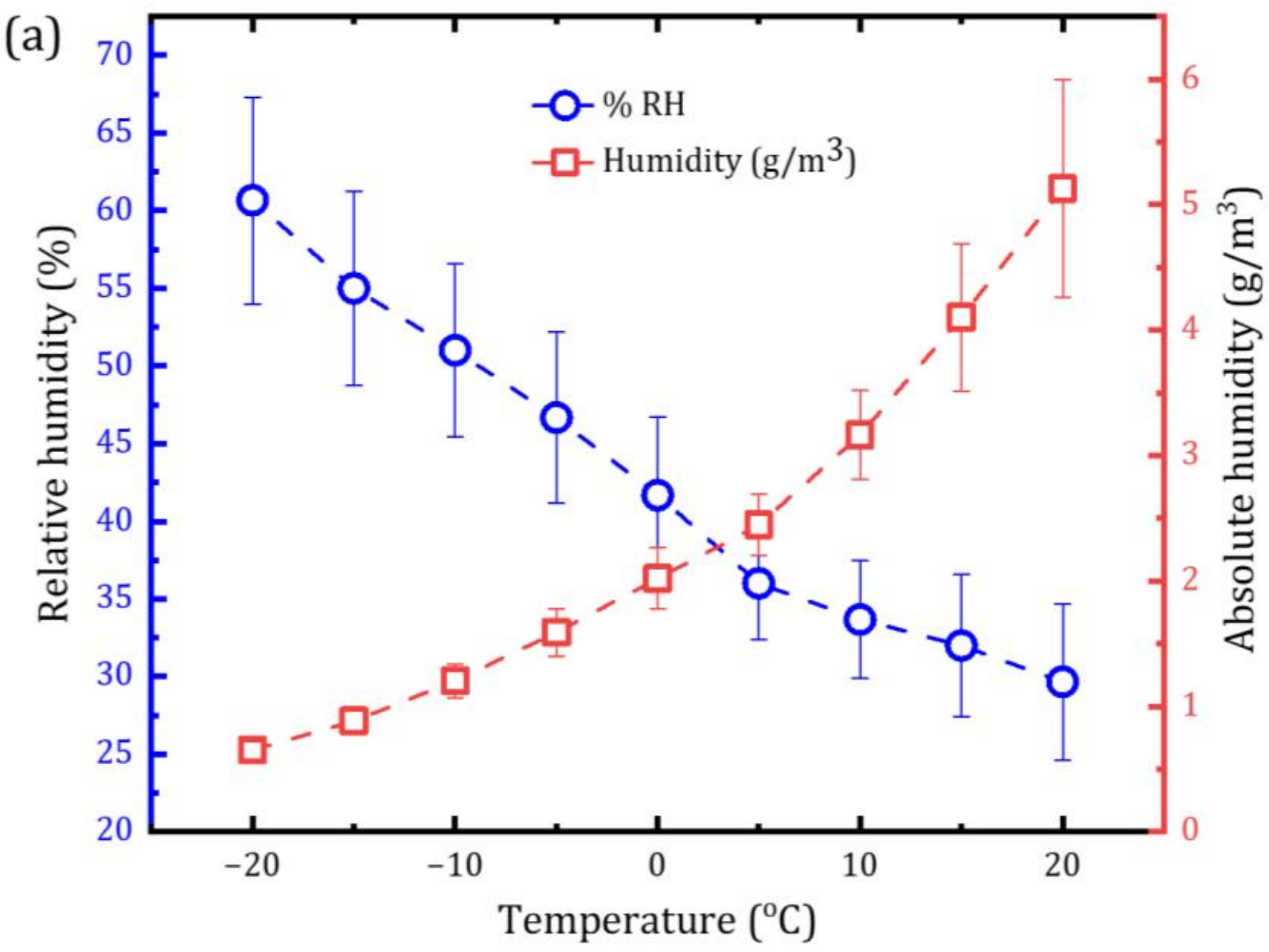

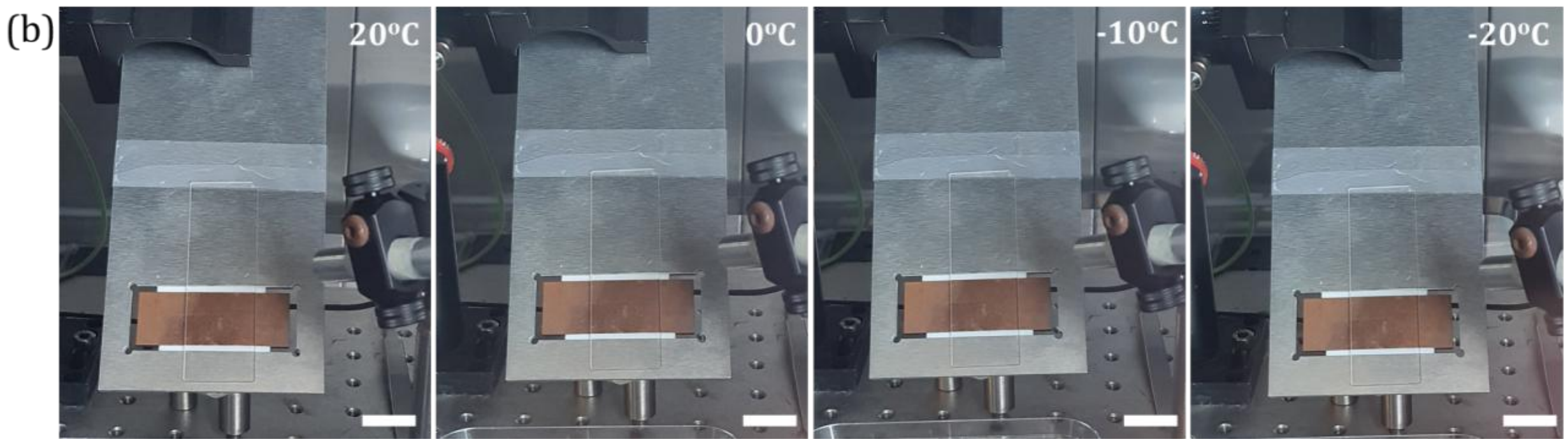

Supplementary Fig. S9: (a) Variation of relative (%RH) and absolute humidity ($g/m^3$) with temperature (±1°C). (b) Photographs of setup at various temperature where no frosting and condensation is observed. Scale = 2 cm.

| **Molecule** | $\chi$ |
|---|---|
| Diiodomethane | 2.45 |
| 1-bromonaphthalene | 2.44 |
| Water | 2.61 |
| Formamide | 2.61 |
| PFOTS | 3.07 |
| OTS | 2.37 |
| PFOMS | 2.98 |
| PEHMA | 2.38 |

Supplementary Table S1: Average Pauling electronegativity ($\chi$) of various molecules used for calculating $\Delta_{en}$.

| **Sample** | **A (C)** | **B (C)** | **reduced $\chi^2$** |
|---|---|---|---|
| OTS | -2.79e-10 ± 9.6e-11 | 1.35e-09 ± 2.2e-10 | 1.13 |
| PFOTS | -3.26e-10 ± 1.7e-10 | 1.49e-09 ± 3.2e-10 | 1.21 |
| PFOMS | -7.28e-11 ± 4.5e-11 | 5.85e-10 ± 8.2e-11 | 3.26 |
| PEHMA | -1.39e-11 ± 1.15e-11 | -8.86e-12 ± 2.1e-11 | 0.04 |
| APTES-PFOTS | 7.91e-12 ± 3.8e-11 | 1.31e-10 ± 8.5e-11 | 0.12 |

Supplementary Table S2: Fitting parameters A, B and reduced $\chi^2$ for function $Q = A \ln(T_m - T) + B$ across all the tested surface. The very small $\chi^2_r$ values for PEHMA and APTES-PFOTS likely reflect conservative uncertainty estimates and the limited number of data points for these substrates, which reduce the discriminatory power of the fit.